\documentclass[a4paper,11pt]{article}
\usepackage{latexsym}
\usepackage{amsmath}
\usepackage{amssymb}
\usepackage{bm}
\title{ Landau gauge condensates from global color model}
\author{$\mbox{Zhao Zhang}^{a,b,1}$, $\mbox{Wei-qin
Zhao}^{a,b,2}$\\[5pt]
\textit{${}^a$CCAST(World Laboratory), P.O. Box 8730,}\\
\textit{Beijing 100080, P. R. China}\\
\textit{${}^b$Institute of High Energy
Physics, Academica Sinica,}\\ \textit{Beijing, 100039, P. R.
China}}
\date{}
\begin{document}
\maketitle
 \footnotetext[1]{E-mail address: zhangzh@ccastb.ccast.ac.cn }
 \footnotetext[2]{E-mail address: chaowq@ccast.ac.cn }
\begin{abstract}
We compute the dimension-2 gluon pair condensate
$g^2\langle{AA}\rangle$ and the dimension-4 mixed quark-gluon
condensate $\langle\overline{q}gA\cdot\gamma{q}\rangle$ in Landau
gauge within the framework of global color model. The result for
the dynamical gluon mass is within the range given by other
independent determinations. The obtained mixed Landau gauge
condensate $\langle\overline{q}gA\cdot\gamma{q}\rangle$ is clearly dependent on
the definitions of the condensates. We show that the consistent
result may be obtained when the same definitions are used.
\\[4pt] {\textit{Keywords}}:\, QCD vacuum ; Global color
 model; Dyson-Schwinger equations; Condensates; Mass gap\\
PACS number(s): 12.38.Lg; 11.15.Tk; 12.38.Aw; 12.39.-x
\end{abstract}
\newpage
\noindent{\textbf{\Large{1. Introduction}}} \vspace{5pt}

The vacuum condensates are believed to play an important role for
characterizing the nonperturbative  aspect of QCD. The well known
gauge invariant condensates such as $\langle\overline{q}q\rangle$
and $\langle\alpha_sG^2\rangle$ are proved to be crucial in the
application of SVZ sum rules. Recently, there has been growing
evidence for the existence of the dimension 2 condensate
$\langle{A^a_\mu}{A^a_\mu}\rangle$ in $SU(N_c)$ pure Yang-Mills
theory and QCD.  Lavelle and Schaden {\cite{r1}} was responsible
for first introducing the non vanishing
$\langle{A^a_\mu}{A^a_\mu}\rangle$ to study the gauge dependent
Green functions by making use of the operator product expansion
(OPE). The recent evidence for the importance of the condensate
$\langle{A^a_\mu}{A^a_\mu}\rangle$ based on OPE can be found in
{\cite{r2,r3,r4}}. The non-vanishing gluon pair condensate gives
rise to the dynamical gluon mass which is determined by
\begin{equation}\label{GM}
m^2_g=\frac{N_c}{(N^2_c-1)D}(\alpha+5-D)\langle{g^2A^a_{\mu}A^a_\mu}\rangle,
\end{equation}
where $\alpha$ stands for the Lorentz covariant gauge fixing
parameter and $D$ is the space-time dimension.

\par
Early implications of the gluon pair condensate have been explored
by Celenza and Shakin {\cite{r5}}. Naively, the operator $A^2$ is
not gauge invariant. The importance of the condensate raises the
question of gauge invariance and leads to a number of papers which
discuss how this condensate could play a role in gauge invariant
formulation {\cite{r6,{r7},{r8},{r9},{r10}}}. Kondo {\cite{r6}}
introduced a BRST invariant and Anti-BRST invariant combination of
dimension 2 operator
\begin{equation}\label{Kondo}
O=\frac{1}{\Omega}{\int}d^Dxtr\left[\frac{1}{2}A_{\mu}(x)A_{\mu}(x)-i{\alpha}C(x)\overline{C}(x)\right],
\end{equation}
where $C(x)$ and $\overline{C}(x)$ are the Faddeev-Popov ghost and
$\Omega$ is the volume of the D-dimensional space-time. The vacuum
expectation value (VEV) of this dimension 2 operator reduces to
the gluon condensate in Landau gauge with $\alpha=0$ which has a
definite physical meaning {\cite{r7}} .
\par
The role of the dimension 4 mixed quark gluon condensate
$\langle{g}\overline{q}\frac{\lambda^a}{2}A^a_{\mu}\lambda^{\mu}q\rangle$
has been studied in {\cite{r11,r12}} in the Lorentz covariant
gauge and it has been shown that this condensate has the
contributions to the quark propagator and the gluon propagator.
This mixed condensate is clearly gauge non-invariant. Recently,
the authors of Ref. {\cite{r4}} have evaluated the values of the
gluon pair condensate and the dimension 4 mixed condensate from
the quenched lattice results for the quark propagator in the
Landau gauge and claimed that it is the first determination of the
mixed condensate
$\langle{g}\overline{q}\frac{\lambda^a}{2}A^a_{\mu}\lambda^{\mu}q\rangle$.
This has motivated us to investigate both of these Landau gauge
condensates from the Global Color Model.

\noindent{\textbf{\Large{2. Formalism}}} \vspace{5pt}

The Global Color Model(GCM) is a quark-gluon filed theory with the
action
\begin{equation}\label{ACT}
S_{GCM}[\overline{q},q,A_\mu^a]=\int(\overline{q}(\gamma\cdot\partial+m-igA_\mu^a\frac{\lambda^a}{2}\gamma_\mu)q+\frac{1}{2}A_\mu^aD^{-1}_{\mu\nu}(i\partial)A_\nu^a)
\end{equation}
and the generating functional
\begin{equation}
Z[J,\overline{\eta},\eta]={\int}D\overline{q}D{q}DAexp({-S_{GCM}[\overline{q},q,A_\mu^a]+\overline{\eta}q+\overline{q}\eta+J_\mu^{a}A_\mu^a}).
\end{equation}
The essence of GCM is that it models the QCD local gluonic action
$\int{F^a_{\mu\nu}}{F^a_{\mu\nu}}$ which has local color symmetry,
by a highly nonlocal action which has a global color symmetry. To
quote Cahill and Gunner {\cite{r13}}: ``There is an Infrared(IR)
saturation effect which in conjunction with the dynamical breaking
of chiral symmetry, appears to suppress details of the formal
color gauge symmetry of QCD." The main aspects of GCM have been
reviewed in {\cite{r13,r14,r15}}.
\par
In Euclidean metric, with ${\{\gamma_\mu,\gamma_\nu\}}=
2\delta_{\mu\nu}$ and $\gamma_\mu^{+}=\gamma_\mu$, the inverse of
the dressed quark propagator in the chiral limit has the
decomposition
\begin{equation}
G^{-1}(p)=i\gamma\cdot{p}+\Sigma(p)=i\gamma\cdot{p}A(p^2)+B(p^2),
\end{equation}
where $\Sigma(p)$ stands for the dressing self-energy of quarks.
Within the GCM formalism, the quark self-energy is determined by
the rainbow-ladder truncated quark DSE
\begin{eqnarray}\label{DSE}
\Sigma(p)=\frac{4}{3}\int\frac{d^4q}{(2\pi)^4}g^2D_{\mu\nu}(p-q)\gamma_{\mu}G(q)\gamma_\nu,
\end{eqnarray}
where $g^2D_{\mu\nu}(p-q)$ is the effective gluon propagator. In
Landau gauge with
$g^2D_{\mu\nu}(p)=(\delta_{\mu\nu}-\frac{p_{\mu}p_{\nu}}{p^2})g^2D(p^2)$,
the nontrivial solution to Eq.(\ref{DSE}) which is characterized
by $B(p^2)\neq{0}$ is determined by the two coupled integrable
functions
\begin{eqnarray}\label{VP}
(A(p^2)-1)p^2&=&\frac{4}{3}\int\frac{d^4q}{(2\pi)^4}g^2D(p-q)(p\cdot{q}+2\frac{q\cdot(p-q)p\cdot(p-q)}{(p-q)^2})\frac{A(q^2)}{q^2A^2(q^2)+B^2(q^2)}\\
\label{SP}B(p^2)&=&4\int\frac{d^4q}{(2\pi)^4}g^2D(p-q)\frac{B(q^2)}{q^2A^2(q^2)+B^2(q^2)},
\end{eqnarray}
while the trivial solution to Eq.(\ref{DSE}) which is
characterized by $B(p^2)=0$ is determined by
\begin{equation}
(A'(p^2)-1)p^2=\frac{4}{3}\int\frac{d^4q}{(2\pi)^4}g^2D(p-q)(p\cdot{q}+2\frac{q\cdot(p-q)p\cdot(p-q)}{(p-q)^2})\frac{1}{q^2A'(q^2)}.
\end{equation}
Within the GCM formalism, the ``Nambu-Goldstone" phase is
described by the nontrivial quark propagator since chiral symmetry
is dynamically broken with the no-vanishing constituent quark mass
$M(p^2)=B(p^2)/A(p^2)$ in the chiral limit and the dressed quarks
are confined since it does not have a Lehmann representation. The
``Wigner"  phase is described by the corresponding trivial quark
propagator $G'(p)=\frac{1}{i\gamma\cdot{p}A'(p^2)}$ with neither
chiral symmetry breaking nor confinement. The quark propagators
$G(x-y)$ and $G'(x-y)$ can be seen as the expectation values of
the operator $Tq(x)\overline{q}(y)$ over the physical vacuum state
$|V\rangle$ and the perturbative vacuum state $|P\rangle$
respectively within GCM formalism.
\par
Since the functional integration over the gluon field $A_\mu^a$ is
quadratic according to (\ref{DSE}), one can perform the
integration over gluon field analytically.  Taking  the quark
color current
$j_\mu^a=ig\overline{q}\frac{\lambda^a}{2}\gamma_\mu{q}$ as the
external source term  for the gluon field $A_\mu^a$, we have the
typical gaussian integrations
\begin{equation}\label{gvev}
\begin{split}
\int\mathcal{D}Ae^{-\frac{1}{2}AD^{-1}A+jA}
&={e}^{\frac{1}{2}jDj}\\
\int\mathcal{D}AAe^{-\frac{1}{2}AD^{-1}A+jA}
&=(jD){e}^{\frac{1}{2}jDj}\\
\int\mathcal{D}AA^2e^{-\frac{1}{2}AD^{-1}A+jA}
&=[D+{(jD)}^2]{e}^{\frac{1}{2}jDj}.
\end{split}
\end{equation}
Through above technique, the gluons vacuum averages are replaced
by the quark color current $j_\mu^a$ together with the effective
gluon propagator $D$. This method is very similar to the
determination of the VEVs of the QCD operators involving the gluon
fields in the instanton liquid model, where by integrating over
the instanton coordinates, one derives an effective quark action
of the form of a Nambu-Jona-Lasinio model {\cite{r16}}. This
method was first used by Meissner {\cite{r17}} to calculate the
gauge invariant dimension 5 mixed quark-gluon condensate
$\langle{g\overline{q}\sigma{G}q}\rangle$ in GCM. In the mean
field level, it is straightforward to calculate the vacuum
expectation value (VEV) of any quark operator of the form
\begin{equation}\label{qop}
\mathcal{O}_n\equiv(\overline{q}_{j_1}\Lambda^{(1)}_{j_1i_1}q_{i_1})
(\overline{q}_{j_2}\Lambda^{(2)}_{j_2i_2}q_{i_2})\cdots
(\overline{q}_{j_n}\Lambda^{(n)}_{j_ni_n}q_{i_n}) ,
\end{equation}
where $\Lambda^{(i)}$ stands for an operator in Dirac, color and
flavor space. The expression for the VEV of the operator
$\mathcal{O}_n$ takes the form
\begin{eqnarray}\label{vev}
\langle\mathcal{O}_n\rangle=(-1)^n\sum_p(-)^p[\Lambda^{(1)}_{j_1i_1}\cdots
\Lambda^{(n)}_{j_ni_n}G_{i_1j_{p(1)}}{\cdots}G_{i_nj_{p(n)}}],
\end{eqnarray}
where $p$ stands for the permutation of the $n$ indexes. According
to (\ref{vev}), one can obtain the familiar expression for the
quark condensate
\begin{equation}\label{qqc}
\langle\overline{q}q\rangle=-\frac{3}{4\pi^2}\int_0^{\infty}dss\frac{B(s)}{sA^2(s)+B^2(s)}.
\end{equation}
The expression for the four quark condensate
$\langle\overline{q}\Lambda^{(1)}q\overline{q}\Lambda^{(2)}q\rangle$
can also be obtained within this formalism which is consistent
with the result based on the vacuum saturation approximation.
\par
Note that there is UV divergence in the calculation of the vacuum
condensates based on Eq.(\ref{vev}). The UV divergence of
Eq.(\ref{vev}) can be traced back to the UV behavior of both the
scalar quark self energy function $B(p^2)$ and the vector quark
self energy function $A(p^2)$. The ultraviolet behavior of the
constituent quark mass $M(p^2)=B(p^2)/A(p^2)$ takes the form in
the Landau gauge\cite{r17.5}
\begin{equation}\label{MUV}
M(p^2)=\frac{2\pi^2\gamma_m}{3}\frac{-\langle\overline{q}q\rangle^0}{p^2(\frac{1}{2}\ln[\frac{p^2}{\Lambda^2_{QCD}}])^{1-\gamma_m}},
\end{equation}
where $\gamma_m=12/(33-2N_f)$ is the mass anomalous dimension,
with $N_f$ the number of light-quark and
$\langle\overline{q}{q}\rangle^0$ the
renormalization-group-invariant condensate. It is easy to see that
the Eq.(\ref{qqc}) is logarithmic divergent according to the UV
behavior of $B(p^2)/A(p^2)$. In Ref {\cite{r18}}, the quark
condensate is well-defined and the UV divergence problem of the
quark condensate is conquered by multiplying a renormalization
constant. Within the framework of GCM,  Eq.(\ref{qqc}) will be
free of UV divergence if the chosen effective gluon propagator has
finite momentum range. Since the vacuum condensates are mainly
determined by the low and medium momentum properties of the
Schwinger functions of QCD, ignoring the effects from the hard
gluonic radiative corrections will have little impact on the
determination of these condensates. Note that the definition of
the vacuum condensate in Ref {\cite{r17}}, which used a finite
cutoff on the integral, is incorrect since the obtained values for
the quark condensate and the mixed quark-gluon condensate are
distinctly sensitive to the cutoff. Without the cutoff, exploring of
the thermal properties of $g\langle{\bar{q}\sigma{G}q}\rangle$ based on GCM
formalism was made in Ref.\cite{r18.1}.

\par
Even in the case of not taking into account the asymptotic
behavior of the effective gluon propagator, the UV behavior of the
function $A(s)$ can also lead to the UV divergence problem in the
calculation of the condensates based on Eq.(\ref{vev}). In Ref
{\cite{r19}}, the calculation of the mixed tensor susceptibility
was performed based on (\ref{vev}) where the UV divergence caused
by the UV behavior of the function $A(s)$ was cancelled by
subtracting the corresponding perturbative contribution which was
evaluated by the trivial quark propagator $G'(p)$. The UV
divergences induced by the function $A(s)$ also appear in the
calculation of the Landau gauge condensates in this letter.
Actually, this kind of divergence is due to no subtraction of the
perturbative contribution in the calculation since in this case
the Eq.(\ref{vev}) is also divergent if one substitutes $G'(x-y)$
for $G(x-y)$.   Because the condensates reflect the
nonperturbative structure of QCD vacuum, the perturbative
contribution to Eq.(\ref{vev}) must be subtracted. In the
framework of GCM, the perturbative contribution can be self
consistently evaluated by replacing the nontrivial quark
propagator $G$ in Eq.(\ref{vev}) with the corresponding trivial
quark propagator $G'$. As a matter of fact, the trivial quark
propagator has been used extensively to play the role of the
perturbative dressed quark propagator in the study of thermal
properties of QCD within the DSE formalism, where the bag constant
was defined as the difference of the pressure between the true QCD
vacuum and the perturbative QCD vacuum, which were evaluated by
the Nambu-Goldstone solution and the Wigner solution to the quark
propagator, respectively\cite{r20}.
\par
Therefore, the formula for the calculation of the VEV of operators
in terms of quark fields should be changed by subtracting the
corresponding trivial contribution
\begin{eqnarray}\label{nvev}
\langle:\mathcal{O}_n:\rangle&=&(-1)^n\sum_p(-)^p[\Lambda^{(1)}_{j_1i_1}\cdots
\Lambda^{(n)}_{j_ni_n}G_{i_1j_{p(1)}}{\cdots}G_{i_nj_{p(n)}}]\nonumber\\
&&-(-1)^n\sum_p(-)^p[\Lambda^{(1)}_{j_1i_1}\cdots
\Lambda^{(n)}_{j_ni_n}G'_{i_1j_{p(1)}}{\cdots}G'_{i_nj_{p(n)}}].
\end{eqnarray}
In fact, for the cases of the determination of the quark
condensate and gauge invariant mixed quark gluon condensate, there
are no needs to subtract these trivial terms which are rigorously
zero due to $B(p^2)\equiv0$.
\par
Since the GCM is not renormalizable and the adopted effective
gluon propagator in our calculation has a finite range in the
momentum space, the scale at which a condensate is defined in our
method is a typical hadronic scale, which is implicitly determined
by the chosen effective gluon propagator $D(s)$ and the
corresponding solution of the quark gap equations. In fact, the
situation is very similar to the calculation of the condensates
based on the instanton liquid model {\cite{r16}}  where the the
scale is set by the inverse instanton size.
\par
\vspace{5pt}
\noindent{\textbf{\Large{3. Evaluation of Landau gauge
condensates}}} \vspace{5pt}
\par
With the above preparation, the gluon pair condensate and the
dimension 4 mixed quark-gluon condensate can be evaluated in the
framework of GCM. Using Eq.(\ref{gvev}) and Eq.(\ref{nvev}) , we
get the condensate $g^2\langle{A^a_{\mu}(0)A^a_{\mu}}(0)\rangle$
\begin{eqnarray}\label{ggAA}
g^2\langle{A_\mu^a(x)A_\mu^a(x)}{\rangle}&=&\langle{g^2D^{aa}_{\mu\mu}}(x-x)\rangle-\int{d^4x'}\int{d^4y'}g^2D^{aa'}
_{\mu\mu'}(x-x')g^2D^{ab'}_{\mu\nu'}(x-y')\nonumber\\
&&\times\langle:{\overline{q}(x')\gamma_{\mu'}\frac{\lambda^{a'}}{2}q(x')\overline{q}(y')\gamma_{\nu'}\frac{\lambda^{b'}}{2}q(y')}:\rangle
\end{eqnarray}
where
\begin{eqnarray}
\langle:{\overline{q}(x')\gamma_{\mu'}\frac{\lambda^{a'}}{2}q(x')\overline{q}(y')\gamma_{\nu'}\frac{\lambda^{b'}}{2}q(y')}:\rangle&=&
tr[\gamma_{\mu'}\frac{\lambda^{a'}}{2}G(x'-y')\gamma_{\nu'}\frac{\lambda^{b'}}{2}G(y'-x')]\nonumber\\
&&-tr[\gamma_{\mu'}\frac{\lambda^{a'}}{2}G'(x'-y')\gamma_{\nu'}\frac{\lambda^{b'}}{2}G'(y'-x')].
\end{eqnarray}
From Eq.(\ref{ggAA}), one can see that the contribution to the
gluon pair condensate is divided into two parts in the mean field
level: the contribution from the effective propagator and the
contribution from the four quark condensate. After Fourier
transformation, we have
\begin{eqnarray}\label{AA}
g^2\langle{A_\mu^a(x)A_\mu^a(x)}{\rangle}&=&24\int{\frac{d^4p}{(2\pi)^4}}g^2D(p^2)+
4\int{\frac{d^4p}{(2\pi)^4}}{\frac{d^4q}{(2\pi)^4}}g^2D(p-q)g^2D(p-q)\nonumber\\
&&\times\bigg\{12B(p^2)B(q^2){Z(p^2)}{Z(q^2)}+\Big[4p\cdot{q}+8\frac{(p-q)\cdot{p}{(p-q)}\cdot{q}}{(p-q)^2}\Big]\nonumber\\
&&\times\Big[A(p^2)A(q^2){Z(p^2)}{Z(q^2)}-A'(p^2)A'(q^2){Z'(p^2)}{Z'(q^2)}\Big]\bigg\}\nonumber\\
&=&\frac{3}{2\pi^2}\int_0^\infty{ds}sg^2D(s)+\frac{1}{8\pi^5}\int_0^\infty{ds}\int_0^\infty{dt}\int_{-1}^{1}{dx}st\sqrt{1-x^2}\nonumber\\
&&\times[g^2D(s+t-2x\sqrt{st})]^2\bigg\{3Z(s)Z(t)B(s)B(t)+\bigg[2\frac{(s-\sqrt{st}x)(\sqrt{st}x-t)}{s+t-2\sqrt{st}x}\nonumber\\
&&+\sqrt{st}x\bigg]\bigg[Z(s)Z(t)A(s)A(t)-Z'(s)Z'(t)A'(s)A'(t)\bigg]\bigg\}
\end{eqnarray}
with $Z(p^2)={1}/({A^2(p^2)p^2+B^2(p^2)})$ and
$Z'(p^2)={1}/({A'^2(p^2)p^2})$.
\par
Applying the same method, we obtain the expression for the mixed
condensate in Landau gauge
\begin{eqnarray}\label{MIX}
-ig\langle\overline{q}A^a_{\mu}\frac{\lambda^a}{2}\gamma_\mu{q}\rangle&=&-4\int{d^4x'}g^2D_{\mu\nu}(x-x')\Big\{
tr_D[\gamma_{\mu}G(x-x')\gamma_{\nu}G(x'-x)]\nonumber\\
&&-
tr_D[\gamma_{\mu}G'(x-x')\gamma_{\nu}G'(x'-x)]\Big\}\nonumber\\
&=&-12\int\frac{d^4p}{(2\pi)^4}\bigg\{\Big[\frac{B^2(p^2)}{A^2(p^2)p^2+B^2(p^2)}+\frac{A(p^2)(A(p^2)-1)p^2}{A^2(p^2)p^2+B^2(p^2)}\Big]\nonumber\\
&&-\Big[\frac{A'(p^2)-1}{A'(p^2)}\Big]\bigg\}\nonumber\\
&=&\frac{3}{4\pi^2}\int_0^\infty{ds}s\left\{\frac{A(s)s}{(A^2(s)s+B^2(s))}-\frac{1}{A'(s)}\right\}.
\end{eqnarray}
Note that the quark DSE
\begin{equation}
\Sigma(x-y)=\frac{4}{3}g^2D_{\mu\nu}(x-y)\gamma_{\mu}G(x-y)\gamma_{\nu}
\end{equation}
has been used again in above derivation, which strongly simplifies
the evaluation of (\ref{MIX}). From (\ref{MIX}), one can see that
the mixed condensate vanishes when QCD undergoes a phase
transition from the Nambu-Goldstone phase to the Wigner phase with
$B\rightarrow{0}$ and $A\rightarrow{A'}$.
\par
As mentioned above, the reduced version of Eqs.(\ref{AA}) and
(\ref{MIX}) without the corresponding subtraction term will be UV
divergent. The UV divergence arises from the behavior of $A(p^2)$
at large momentum region. In the GCM formalism, $A(p^2)$ decreases
with increasing $p^2$ until it takes the value $A(p^2)=1$ at
$p^2=\infty$. This property can be illustrated by a simple
analytic confining infrared-dominant (ID) model
\begin{equation}\label{ID}
g^2D_{\mu\nu}(p-q)=(2\pi)^4\frac{\eta^2}{4}\delta^4(p-q)(\delta_{\mu\nu}-\frac{p_{\mu}p_\nu}{p^2})
\end{equation}
which was proposed in Ref {\cite{r21}}. The ID model has been used
extensively in the literature, which can exhibit many of the
qualitatively features based on more complicated ansantz. The
Nambu-Goldstone solution for this model is
\begin{eqnarray}\label{NGA}
A(p^2)&=&\left\{
       \begin{array}{cc}
             2, &{ \quad p^2<\frac{\eta^2}{4}},\\
              \frac{1}{2}(1+\sqrt{1+\frac{2\eta^2}{p^2}}),
               &{\quad \mbox{otherwise}},
       \end{array}\right.\\
\label{NGB}B(p^2)&=&\left\{
       \begin{array}{cc}
             \sqrt{\eta^2-4p^2}, & {\quad p^2<\frac{\eta^2}{4}}, \\
            0, &{\quad
             \mbox{otherwise}}.
       \end{array}\right.
\end{eqnarray}
The corresponding Wigner solution takes the form
\begin{equation}\label{WIG}
B'(p^2)\equiv0,\quad
A'(p^2)=\frac{1}{2}(1+\sqrt{1+\frac{2\eta^2}{p^2}}).
\end{equation}
We can see that the function $A(p^2)$ is identical to $A'(p^2)$ in
the $p^2>\frac{\eta^2}{4}$ region and monotonously decreases with
$p^2$ until it reaches $1$ at $\infty$. Using the ID model,  the
$A(p^2)$ induced UV divergence without subtraction can be easily
shown. For example, due to $A(s)-1\sim\frac{1}{s}$ for
$s\rightarrow\infty$ according to (\ref{NGA}), the second term of
right hand side of Eq.(\ref{MIX}) is clearly divergent if we do
not adopt the subtraction procedure. From Eq.(\ref{NGA}) and
Eq.(\ref{WIG}), it is also shown that the contribution of the last
term in Eq.(\ref{MIX}) to the mixed condensate is mainly
determined by the low energy  behavior of  the $A(p^2)$ and
$A'(p^2)$.
\par
The chosen model ansatz for $g^2D(s)$ in this letter is the
Gaussian type model
\begin{equation}\label{GAUSSIN}
g^2D^{ab}_{\mu\nu}(q)=4\pi^2\Delta\frac{q^2}{\omega^2}\exp(-\frac{q^2}{\omega^2})\delta^{ab}t_{\mu\nu}(q)
\end{equation}
where the parameter $\Delta$ sets the strength of the interaction
and  $\omega$ is the range parameter at which the scalar function
in the parameterization is maximal. This Gaussian type form was
supposed to represent a sensible hadron model in \cite{r22} and
the parameter $\omega$ was envisaged to have a value of several
hundred MeV which sets the interaction scale. In our work, these
parameters are constrained by giving the reasonable constituent
quark mass $M_q(q^2=0)$ and the pion decay constant fixed at $87$
MeV which is more appropriate in the chiral limit rather than the
Pion's mass-shell value of $93$ MeV. To check the sensitivity of
our results on the model effective gluon propagator, various sets
of parameters $\Delta$ and $\omega$ were used in our calculations.

\vspace{10pt}
\begin{tabular}{cccccccc}
\multicolumn{8}{l}{Table 1. Calculated Landau gauge condensates
 from the Gaussian model in the chiral }\\
\multicolumn{8}{l}{limit with three sets of parameters. The quark
condensate and the constituent quark }\\
\multicolumn{8}{l}{mass from this model are also listed. The units for $\Delta$, $\omega$, $M_g$, $M_q$ and $f_\pi$ are GeV.}\\[2pt]\hline\hline
$\Delta$&$\omega$&$\langle\overline{q}q\rangle(\mbox{GeV}^3)$
&$g^2\langle{AA}\rangle(\mbox{GeV}^2)$&$g\langle{\overline{q}A\cdot\gamma{q}}\rangle(\mbox{GeV}^4)$&$M_g$&$M_q(0)$&$f_\pi$
\\[2pt]\hline
30.0&0.40&$-0.215^3$&$9.5$&$-1.2*10^{-3}$&0.94&$0.381$&0.087\\\hline
19.0&0.45&$-0.223^3$&$9.2$&$-1.3*10^{-3}$&0.93&$0.342$&0.087\\\hline
13.2&0.50&$-0.232^3$&$9.9$&$-1.4*10^{-3}$&0.96&$0.311$&0.087\\\hline\hline
\end{tabular}
\vspace{10pt}
\par
Table 1 shows that the dynamical gluon mass obtained in the
Gaussian model is around $1\mbox{GeV}$. This value is compatible
with the estimates obtained by other approaches, which are within
the range $0.5-1.5\mbox{GeV}$ {\cite{r23}}. Actually, from our
numerical studies, the main contribution to the gluon pair
condensate comes from  the first term of right hand side of
Eq.(\ref{ggAA}). This is shown in Table 2 with the dynamical gluon
masses $M_{g1}$ which is determined only from the first term of
the Eq.(\ref{ggAA}). Comparing $M_{g1}$ with $M_g$, we get the
conclusion that the gluon pair condensate is mainly determined by
the low energy part of $g^2D(p)$ and the correction from the four
quark condensates can be neglected. This is reasonable since in
GCM the $g^2D(p)$ is interpreted as the effective gluon
propagator. Our results further confirms that the Gaussian model
 is surely an effective quark-quark interaction form in the low
momentum region. In addition, this also provides a constraint to
the form of $g^2D(p)$ in GCM, which should give a reasonable
dynamical gluon mass.
\par
 Neglecting the four quark condensate contribution, the gluon pair condensate
 from the ID model takes the simple form
\begin{equation}\label{IDGPC}
 g^2\langle{A^2}\rangle=2N_c\eta^2.
\end{equation}
   In the case
 $\eta^2/2=m_\rho^2=0.59\mbox{GeV}^2${\cite{r21}}, the gluon pair condensate takes the value $7.1\mbox{GeV}^2$
 and the corresponding gluon mass is $0.82\mbox{GeV}$ which is also in agreement with the estimates
 from other methods.

\par
In Table 1, the magnitude of the obtained dimension 4 mixed
condensate is much smaller in comparison with the recent results
based on the quenched lattice results for the quark propagator in
Landau gauge\cite{r4}. Within the ID model with
$\eta^2/2=0.59\mbox{GeV}^2$, the mixed condensate takes the
value $-0.0003\mbox{GeV}^4$. However, the magnitude of the quark
condensate $-\langle\overline{q}q\rangle^{1/3}$ obtained in ID
model with the same parameter is $118\mbox{MeV}$ which is also
smaller in comparison with the standard value $250\mbox{MeV}$. If
we simply lift the the value of $\eta$, for example, taking
$\eta=1.8\mbox{GeV}$, we get that
$g\langle{\overline{q}A\cdot\gamma{q}}\rangle=-0.0018\mbox{GeV}^4$
with $-\langle\overline{q}q\rangle^{1/3}=184\mbox{MeV}$.
Therefore,  the mixed condensates obtained in both the ID and Gaussian models
 are at the same order of magnitude.
According to {\cite{r1,r12,r4}, the coefficient of $1/Q^4$ in the
expansion of the vector quark self-energy function at sufficiently
large values of $Q$ is determined by the combined dimension 4
condensates
\begin{equation}\label{MIXC}
\alpha_s\langle\overline{q}gA\cdot\gamma{q}\rangle-\frac{4\pi}{27}\langle\frac{\alpha_s}{\pi}G^2\rangle,
\end{equation}
where the gluon condensate takes the value
$\langle\frac{\alpha_s}{\pi}G^2\rangle\simeq0.01\mbox{GeV}^4$.
Multiplying $\alpha_s(\mu)\simeq{0.5}$ to the mixed condensate
where $\mu$ stands for a typical hadronic mass, we find that
$\alpha_s\langle\overline{q}gA\cdot\gamma{q}\rangle$ is around
$-0.0007\mbox{GeV}^4$ in our calculation. In comparison with the
magnitude of $\langle\frac{\alpha_s}{\pi}G^2\rangle$, we find that
the main contribution to Eq.(\ref{MIXC}) is determined by the
gluon condensate. This result is in contrary to the one obtained in
Ref {\cite{r4}}. The obtained value for
$\alpha_s\langle\overline{q}gA\cdot\gamma{q}\rangle$  in  Ref
{\cite{r4}} is $(-0.11\pm0.03)\mbox{GeV}^4$,  comparing to which the
gluon condensate contribution to (\ref{MIXC}) can be neglected.

\vspace{10pt}
\begin{tabular}{ccccc}
\multicolumn{5}{l}{Table 2. The comparison of the contributions to
the gluon pair
condensate from }\\
\multicolumn{5}{l}{the effective gluon propagator and the four quark condensate in Eq.(\ref{ggAA}).}\\[2pt]\hline\hline
$\Delta(\mbox{GeV})$&$\omega(\mbox{GeV})$&$g^2\langle{AA}\rangle_1(\mbox{GeV}^2)$&$g^2\langle{AA}\rangle_2(\mbox{GeV}^2)$&$M_{g1}(\mbox{GeV})$
\\[2pt]\hline
30.0&0.40&$9.22$&$0.28$&0.93\\\hline
19.0&0.45&$9.35$&$-0.13$&0.94\\\hline
13.2&0.50&$9.90$&$-0.05$&0.96\\\hline\hline
\end{tabular}
\vspace{10pt}
\par
According to the quark motion equation
$(\gamma_{\mu}\partial_{\mu}+m-i\frac{\lambda^a}{2}A^a_{\mu}\gamma_{\mu}){q}=0$
 , we expect that the condensate
$\langle\overline{q}(\gamma_{\mu}\partial_{\mu}+m-i\frac{\lambda^a}{2}A^a_{\mu}\gamma_{\mu})q\rangle$
should take zero. In the chiral limit, this relation is reduced to
\begin{equation}\label{QME}
-i\langle\overline{q}\frac{\lambda^a}{2}A^a_{\mu}\gamma_{\mu}q\rangle=-\langle\overline{q}\gamma_{\mu}\partial_{\mu}q\rangle.
\end{equation}
Actually, the evaluation of the Landau gauge condensate
$\langle\overline{q}\gamma_{\mu}\partial{\mu}q\rangle$ is very
simple in GCM. Repeating the procedure above, including the subtraction, we get
\begin{eqnarray}\label{QBPQ}
-\langle\overline{q}(x)\gamma_{\mu}\partial_{\mu}q(x)\rangle&=&\partial^{x'}_{\mu}\left\{tr[G(x'-x)\gamma_{\mu}]
-tr[G'(x'-x)\gamma_{\mu}]\right\}|_{x'={x}}\nonumber\\
&=&12\int\frac{d^4p}{(2\pi)^4}\left\{\left[\frac{A(p^2)p^2}{A^2(p^2)p^2+B^2(p^2)}\right]-\left[\frac{1}{A'(p^2)}\right]\right\}\nonumber\\
&=&\frac{3}{4\pi^2}\int_0^\infty{ds}s\left[\frac{A(s)s}{A^2(s)s+B^2(s)}-\frac{1}{A'(s)}\right].
\end{eqnarray}
It can been see that this is just the Eq.{(\ref{MIX})}. However,
without the subtraction, Eq.(\ref{MIX}) reduces to
\begin{equation}
-\frac{3}{4\pi^2}\int_0^\infty{ds}s\left[\frac{B^2(s)}{A^2(s)s+B^2(s)}+\frac{A(s)(A(s)-1)s}{A^2(s)s+B^2(s)}\right],
\end{equation}
which is negative and divergent since $A(s)>1$ for finite $s$,
while Eq.(\ref{QBPQ}) reduces to
\begin{equation}
\frac{3}{4\pi^2}\int_0^\infty{ds}s\left[\frac{A(s)s}{A^2(s)s+B^2(s)}\right],
\end{equation}
which is positive and divergent. Therefore, there exists a large
difference between these two expressions even in sign if we do not
adopt the subtraction mechanism.
This shows that it is reasonable to introduce the subtraction mechanism in our definition.
For the massive case, according
to the PCAC relation, the value of $m\langle\overline{q}q\rangle$
is around $-0.00008\mbox{GeV}^4$, which suggests that the Landau
gauge dimension 4 mixed condensate is an order of magnitude larger
than the $m\langle\overline{q}q\rangle$ condensate.
\par
Here we still want to stress that the discrepancy of the estimates of the mixed quark gluon condensate
arises from the different definitions of condensates. In Ref\cite{r4} and those related to it,
the condensates are defined as the coefficients of expansion terms in A(s). Actually, by taking the same
definition,  similar results can also be obtained within Global Color Model. Consider
the ID model described by Eqs.(\ref{NGA}) and (\ref{NGB})
\begin{equation}
    A(s)=\frac{1}{2}\Big[1+\sqrt{1+2\eta^2/s}\Big]\overset{large-s}{=}1+\frac{\eta^2}{2}\frac{1}{s}-\Big[\frac{\eta^2}{2}\Big]^2\frac{1}{s^2}+\cdots
\end{equation}
With the conventions of Ref.\cite{r4} one would read
\begin{eqnarray}\label{AEPID1}
 \alpha_s\langle{A^2}\rangle=\frac{N_c}{\pi}\frac{\eta^2}{2},\\
 \label{AEPID2}   \alpha_s\langle{\bar{q}gA\cdot\gamma{q}}\rangle-\frac{4}{9N_c}\langle{\frac{\alpha_s}{\pi}}G^2\rangle
    =\frac{4}{3\pi}\Big[\frac{\eta^2}{2}\Big]^2.
\end{eqnarray}
Note that Eq.(\ref{AEPID1}) is identical to Eq.(\ref{IDGPC}). Substituting $\eta^2/2=0.59\mbox{GeV}^2$ into Eqs.(\ref{AEPID1}) and (\ref{AEPID2}) one gets
\begin{eqnarray}
 \alpha_s\langle{A^2}\rangle=0.57\quad\mbox{GeV}^2 ,\\
   \alpha_s\langle{\bar{q}gA\cdot\gamma{q}}\rangle-\frac{4}{9N_c}\langle{\frac{\alpha_s}{\pi}}G^2\rangle
    = -0.15\quad\mbox{GeV}^2,
\end{eqnarray}
which are in agreement with the results of Ref.\cite{r4}.
\par
The small value of the mixed quark gluon condensate obtained based on the definition (\ref{vev})
 can be tracked back to subtracting the coefficients $1/s$ and $1/s^2$ in the expansion of $A'(s)$
 evaluated in the Wigner phase from those obtained with $A(s)$ calculated in the Nambu phase. In
  the case of sufficiently large value of s
 \begin{equation}
       A(s)=1+c_1^N\frac{1}{s}+c_2^N\frac{1}{s^2}\quad,\quad A'(s)=1+c_1^W\frac{1}{s}+c_2^W\frac{1}{s^2},
\end{equation}
one gets
\begin{equation}\label{diffcon}
\alpha_s\langle{\bar{q}gA\cdot\gamma{q}}\rangle-\frac{4}{9N_c}\langle{\frac{\alpha_s}{\pi}}G^2\rangle=c_2^N-c_2^W.
\end{equation}
The fact that the vector self-energy function of the quark propagator is not too sensitive to the phase in large momentum, i.e., $c_2^N\approx{c_2^W}$,
can qualitatively explain why the l.h.s of Eq.(\ref{diffcon}) is small according to our definition.

\vspace{5pt} \noindent{\textbf{\Large{4.Discussion and conclusion
}}} \vspace{5pt}
\par
We have evaluated the gluon pair condensate and the dimension 4
mixed quark gluon condensate in Landau Gauge within the framework
of GCM. To avoid divergence and subtract the perturbative contribution,
two measures are taken in our calculations.
The input effective gluon propagator adopted in our calculation
has finite range in momentum space, which can suppress the divergence induced by the UV behavior of the scalar
quark self-energy function. This is reasonable since the condensates are mainly determined by the low energy
properties of QCD. To subtract the perturbative
contribution, the trivial solution or the Wigner solution to the
quark DSE was used self-consistently in our formalism.
\par
With above method, the contributions to the gluon pair condensate was decomposed into
two parts: the contribution from the effective gluon propagator
and the contribution from the four quark condensate. From our numerical study, we find that the
correction to the gluon pair condensate from the four quark condensate is very small and can be neglected. The obtained
gluon pair condensate and the corresponding dynamical
gluon mass are compatible with other estimations in the
literatures. The expression for the dimension 4 mixed quark gluon
condensate is very simple in our formalism. This expression has
been verified by using the quark motion equation. The value of
the mixed condensate $\langle{g\overline{q}A\cdot\gamma{q}}\rangle$ is an order of magnitude larger than the
condensate $m\langle\overline{q}q\rangle$.

\par
We also explicated that the discrepancy of the estimates of the Landau mixed quark gluon condensate between our formalism and
others arises from the different definitions  of the condensates. If we take the same definition as in Ref.\cite{r4},
the obtained values of both the gluon pair condensate and the mixed condensate in GCM are consistent with the
corresponding ones extracted from Lattice simulation results. That means when the same definitions are used, the GCM/DSE
formalism is completely consistent with the analysis based on the lattice results.
Note that due to the ambiguity in the definitions of the condensates, there exists no natural or unique definitions but only
clear definitions; when the same definitions are used, all well-constrained models should yield the same result. This fact is made very clear
for the quark condensate $\langle{\bar{q}q}\rangle$ \cite{r24}.

\end{document}